\documentclass[aps,prl,twocolumn,showpacs,amsmath,amssymb,groupedaddress]{revtex4}

\usepackage{graphicx} 
\usepackage{dcolumn}
\usepackage{bm}
\usepackage{epstopdf}

\usepackage{times}
\begin{document}

\title{Generating Unexpected Spin Echoes in Dipolar Solids with $\pi$ Pulses}


\author{Dale Li, A. E. Dementyev, Yanqun Dong, R. G. Ramos, and S. E. Barrett}
\email[e-mail: ]{sean.barrett@yale.edu}
\affiliation{Department of Physics, Yale University, New Haven, Connecticut 06511}
\homepage[web: ]{http://opnmr.physics.yale.edu}



\date{\today}

\begin{abstract} 
NMR spin echo measurements of $^{13}$C in C$_{60}$, $^{89}$Y in Y$_{2}$O$_{3}$, and $^{29}$Si in silicon 
are shown to defy conventional expectations when more than one $\pi$ pulse is used.
Multiple $\pi$-pulse echo trains may either freeze out or accelerate the 
decay of the signal, depending on the $\pi$-pulse phase.  Average Hamiltonian theory, combined with exact quantum calculations, reveals an
intrinsic cause for these coherent phenomena: the dipolar coupling has a many-body effect during any real, finite pulse.  
\end{abstract}

\pacs{03.65.Yz, 03.67.Lx, 76.20.+q, 76.60.Lz}

\maketitle


Nuclear magnetic resonance experiments rest upon a solid theoretical foundation \cite{slichter, abragam, mehringText, ernstText}. 
Advanced pulse sequences have been successfully applied
in all subfields of magnetic resonance, in atomic physics, and even in the emerging field of quantum
information processing.
Thus, we were surprised to discover that simple experiments on doped silicon appeared to be inconsistent with 
conventional NMR theory \cite{siYale}.  For example, coherent signals may be observed well beyond the $T_2$ that is measured in two-pulse
spin echo \cite{hahn} experiments, provided that more than one $\pi$ pulse is used \cite{siYale, siJapan, siStanford}.

In this Letter, we report the same surprising phenomena in Buckminsterfullerene (C$_{60}$) and Yttria (Y$_{2}$O$_{3}$), two solids linked to
silicon through the form of the homonuclear dipolar coupling \cite{slichter}.  
We also show that multiple $\pi$-pulse echo trains may either freeze out or accelerate the 
expected dipolar decay of the NMR signal, depending upon the phases used for the $\pi$ pulses.
Average Hamiltonian theory \cite{AHT}, combined with exact quantum calculations, is used to show that this pulse phase sensitivity has an intrinsic origin, 
arising from the surprisingly non-negligible effects of the dipolar coupling during strong but finite pulses.

\begin{figure}[b]
\includegraphics[width=3.4 in]{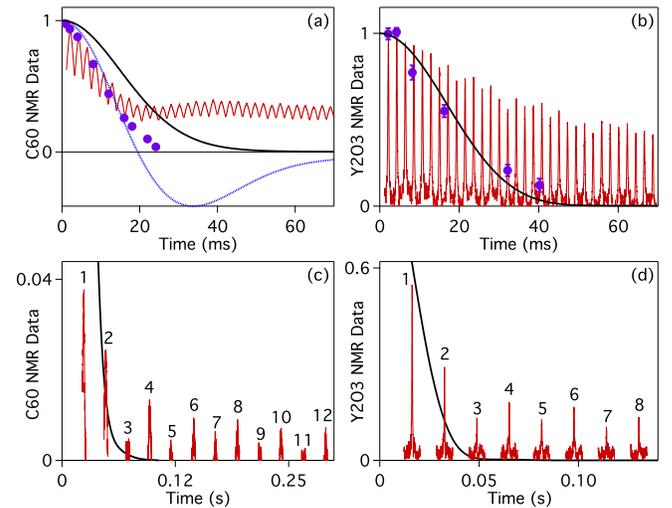}
\caption{\label{fig1}(color online). 
NMR measurements of $^{13}$C  in C$_{60}$ (a), (c) and $^{89}$Y in Y$_2$O$_3$ (b), (d). Experimental parameters: $T\!\!=\!\!300$ K, $B\!\!=\!\!12$ Tesla, 
spin-lattice relaxation time $T_1\!\!=\!\! 26$ s (C$_{60}$), 2.3 hr (Y$_2$O$_3$).  
The dots in (a) and (b) are Hahn echo peaks while the red lines are CPMG echo trains.  Simulated dipolar decay curves are shown for comparison (black and blue, see text).  Note the long tail for short interpulse spacings (a) and (b) and the even-odd effect for long interpulse spacings (c), (d).
}
\end{figure}

The decay of signals produced by the single-$\pi$-pulse Hahn echo sequence \cite{hahn}
(HE: $90_{X}$-$\tau$-$180_{Y}$-$\tau$-ECHO) as 
$\tau$ is increased is a standard measure of $T_2$ \cite{slichter}.  In
both C$_{60}$ [Fig.\ \ref{fig1}(a)] and Y$_{2}$O$_{3}$ [Fig.\ \ref{fig1}(b)] powders,  the
multiple-$\pi$-pulse Carr-Purcell-Meiboom-Gill sequence \cite{cpmg} (CPMG:  
$90_{X}$-\big\{$\tau$-$180_{Y}$-$\tau$-ECHO\big\}$^{\mathrm{repeat}}$) generates
echoes well beyond $T_2$. 
Moreover, CPMG in these samples show both the long tail at short $\tau$ [Fig.\ \ref{fig1}(a) and \ref{fig1}(b)] 
and the even-odd effect at long $\tau$ [Fig.\ \ref{fig1}(c) and \ref{fig1}(d)], as previously reported in silicon \cite{siYale, siJapan}.

At room temperature, C$_{60}$ molecules form an fcc lattice, and each C$_{60}$ undergoes rapid isotropic rotation about its lattice point \cite{spinningC60}.  
The time-averaged spin Hamiltonian is obtained by placing $^{13}$C [spin $I=1/2$, with 1.11\% natural abundance (n.a.)] spins on a 60-fold degenerate fcc lattice.  
Multiple occupancy of an fcc lattice site is not a problem in this model, since the intra-C$_{60}$ dipolar couplings
are averaged to zero by the rapid buckyball rotation.  This motion also eliminates any inter-C$_{60}$ $J$ coupling \cite{slichter} but leaves the dipolar coupling between spins on different buckyballs.  Fig.\ \ref{fig1}(a) and \ref{fig1}(c) are the strongest evidence yet that pure dipolar coupling is sufficient to produce the phenomena \cite{levstein}.  Moreover, dilution of the spins is not required since $^{89}$Y ($I=1/2$, 100\% n.a.) experiments in Y$_{2}$O$_{3}$ [Fig.\ \ref{fig1}(b) and \ref{fig1}(d)]
also look quite similar to the earlier $^{29}$Si ($I=1/2$, 4.67\% n.a.) experiments in doped silicon \cite{siYale, siJapan}.

The NMR signal in both Hahn echo and CPMG experiments is 
 $\propto\!\! \langle I_{y_T} (t)\rangle\equiv  \!\sum_{i}^{N}\!\mathrm{Tr} \{\rho(t)I_{y_{i}}\}$ in the rotating frame \cite{slichter, abragam, mehringText, ernstText}.
The time-dependent density matrix $\rho(t)$ is calculated by starting with its conventional equilibrium value
$\rho(0) \!\!=\!\!\sum_{i}^{N}I_{z_{i}}\!\!=\!\!I_{z_{T}}$,
which assumes both the strong field and the high temperature approximations.
Treating a strong $90_{X}$ pulse as a  perfect $\frac {\pi}{2}$ rotation about $X$, $\rho(0)$ becomes  
$\rho(0^{+}) = e^{i (\pi/2)I_{x_{T}}} \rho(0) e^{-i (\pi/2)I_{x_{T}}} = I_{y_{T}}$.

In between pulses, the spin Hamiltonian for these samples has two main parts.  
The first is the secular part of the dipolar coupling:
\begin{equation}
{\mathcal H}_{zz} \equiv \!\!\! \sum_{j>i}^{N}\! B_{ij}\Big(3I_{z_{i}}I_{z_{j}}-\vec{I}_{i}\cdot\vec{I}_{j}\Big),
\label{Hzz}
\end{equation} 
where  $B_{ij}=\frac{1}{2}\frac{\gamma^{2}\hbar^{2}}{r_{ij}^{3}} [1-3\cos^{2}\theta_{ij}]$, $\gamma$ is the gyromagnetic
ratio, and  ${\vec r_{ij}}$, the vector between spins $i$ and $j$, satisfies
${\vec r_{ij}}\cdot{\hat z} =  r_{ij}\cos\theta_{ij}$ (a static lab field points along $\hat z$).  
The second part is a Zeeman term, ${\mathcal H}_{\Omega_{z_{i}}}\!\!=\!\!\!\sum_{i}^{N}\!\! \! -\hbar\omega_{z_{i}}I_{z_{i}}$,
where $\omega_{z_{i}}$ is the angular frequency offset for spin $i$ relative to on-resonance spins \cite{slichter, abragam, mehringText, ernstText}.
We further simplify this part by dropping the index i, since the linewidth of spectra studied here suggest that $\omega_{z_{i}}$ is extremely uniform and entirely due to bulk diamagnetism of the powder
[e.g., at B=12 Tesla, the linewidth $\Gamma$= 260 Hz (C$_{60}$), 180 Hz (Si:Sb), and 3.1 kHz (Y$_2$O$_3$)].
The spin Hamiltonian describing free evolution is thus
\begin{equation}
{\mathcal H}_{0}={\mathcal H}_{zz}+{\mathcal H}_{\Omega_{z}}={\mathcal H}_{zz}+\Omega_{z}I_{z_{T}}.
\label{H0}
\end{equation}

Two different approximations are used to calculate the expected echo decays shown in Fig. \ref{fig1}.  Both average over many disorder
realizations (DRs), where each DR uses a randomly oriented lattice, with sites randomly occupied to match the natural abundance.  
In the first approximation, we build a closed quantum spin system around a central spin and keep all the terms in Eq. (\ref{Hzz}) while setting $\Omega_{z}=0$.
Calculation of the evolution of the central spin is exact using numerical diagonalization [blue curve, Fig. \ref{fig1}(a), with $N=9$ and 18 DRs].  
Unfortunately, computer limitations make it impractical to include $N>9$ spins in this approach, so it fails for large or dense spin systems.
In the second approximation, we drop the flip-flop terms in Eq. (\ref{Hzz}), 
which yields an analytic expression \cite{siYale} for any $N$ (black curves, Fig. \ref{fig1}, with $N \gg 1000$ and $\gg 150$ DRs). 
This second approach is unjustified for such clean samples, but Y$_2$O$_3$ is beyond the limits of the first method.
Figure \ref{fig1} shows Hahn echo data consistent with the expected decay due to the dipolar coupling.  
However, the CPMG echoes are observed well beyond this limit.

The conventional $\delta$-function pulse approximation \cite{slichter, abragam, mehringText, ernstText} 
treats very strong pulses as instantaneous ${\pi}$ rotations. In this limit, the density matrix for the Hahn echo (SE1) must agree with that for the $n$th Spin Echo (SE$n$) of a CPMG experiment.
Thus the difference between the CPMG and Hahn echo data (Fig.\ \ref{fig1}) is surprising \cite{siYale}.

\begin{figure}[b]
\includegraphics[width=3.4 in]{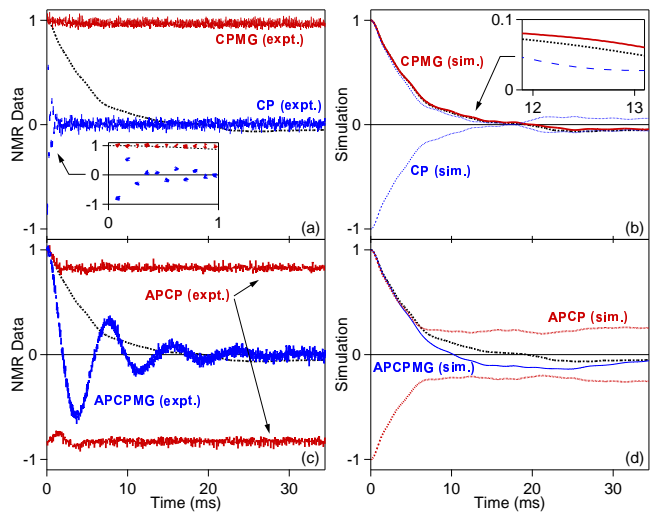}
\caption{\label{fig2}(color online).
(a), (c) $^{29}$Si measurements in Si:Sb ($10^{17}$ Sb$/$cm$^{3}$) with the four phase choices from Table \ref{tab:table1}.  
(b), (d)  $N\!\!=\!\!7$ simulations for the same experimental conditions: $t_{p}\!\!=\!\!14$ $\mu$s, $\tau\!\!=\!\!36$ $\mu$s, $T\!\!=\!\!300$ K, and $B\!\!=\!\!11.75$ Tesla 
($\frac{\gamma \mathrm{B}}{2\pi}\!\!=\!\!99.5$ MHz, $\frac{\omega_1}{2 \pi}\!\!=\!\!35.7$ kHz). 
Each curve is averaged over 1000 DRs with $\Omega_{z}/h$  drawn from a 290 Hz wide gaussian and typical couplings $|B_{12}|/h\!\approx\!44.5$ Hz, $|B_{17}|/h\! \approx\! 3.5$ Hz.  
Inset (a) shows rapid decay of CP data.  Inset (b) shows a distinction between CP and CPMG simulations.
The black dashed reference (a)-(d) sets $\Omega_z = 0$ and turns off ${\mathcal H}_{zz}$ during each pulse.}
\end{figure}

Moreover, the experiments defined in Table \ref{tab:table1} should produce identical $|\langle I_{y_T}(t)\rangle|$ if the $\pi$ pulses are instantaneous.
However,  the measured echo trains for short $\tau$ [Figs.\ \ref{fig2}(a) and \ref{fig2}(c)] exhibit a striking pulse sequence sensitivity (PSS).
In this limit, the measured signal can either stay very close to one (CPMG, APCP), or dive rapidly towards zero (CP, APCPMG).
The PSS is observed \cite{RelExpt} even when BB1 composite $\pi$ pulses \cite{BB1} are used.  

It is natural to attempt to blame the discrepancies between theory and experiment on extrinsic imperfections of instantaneous $180_{Y}$ pulses.  
Examples include misadjustment of the rotation angle, rf inhomogeneities,
and pulse phase transients \cite{slichter,mehringText}.  
We investigated these and other extrinsic errors of instantaneous pulses, but despite experimental improvements \cite{RelExpt}, the effect persists. 
This led us to consider the limits of the conventional $\delta$-function pulse approximation.

\begin{table}[t]
\caption{\label{tab:table1}$\pi$ pulse and echo phases for four pulse sequences of the form:
$90_{X}$-\big\{$\tau$-$180_{\phi_{1}}$-$\tau$-SE1-$\tau$-$180_{\phi_{2}}$-$\tau$-SE2-\big\}$^{\mathrm{repeat}}$.\\}
\begin{ruledtabular}
\begin{tabular}{ccccc}
Sequence & $\phi_{1}$& $\phi_{2}$&SE1&SE2\\
\hline
CP& $+X$ &  $+X$ &  $-Y$ &  $+Y$ \\
APCP& $-X$ &  $+X$ &  $-Y$ &  $+Y$ \\
CPMG& $+Y$ &  $+Y$ &  $+Y$ &  $+Y$ \\
APCPMG& $-Y$ &  $+Y$ &  $+Y$ &  $+Y$ \\
\end{tabular}
\end{ruledtabular}
\end{table}

The full Hamiltonian during an ideal pulse along the $\phi_{i}$ direction is
\begin{equation}
{\mathcal H}_{P_{\phi_{i}}}\!\!\!=\!-\hbar\omega_{1}I_{{\phi_{i}}_{T}}\!+\!{\mathcal H}_{0}\!
\label{Hp}
\end{equation} 
where the pulse's angular frequency, $\omega_{1}$, is the same for all spins. 
The $\delta$-function pulse approximation produces an instantaneous $\pi$ rotation about axis $\phi_{i}$ by keeping $\omega_{1} t_{p}=\pi$ while $\omega_{1}\rightarrow\infty$  and  $t_{p} \rightarrow 0$,  where $t_{p}$ is the pulse duration.

In any real experiment, $t_{p}>0$, so the ${\mathcal H}_{0}$ term in Eq. (\ref{Hp})
might have some non-negligible effect.  This possibility was first raised by Dobrovitski's simulations \cite{slava} of our silicon data \cite{siYale}.  
This is an intrinsic deviation from the instantaneous pulse limit, which cannot be avoided for $t_p>0$, as long as $\mathcal H_0$ is nonzero during the pulse.
In fact, this is the sole cause of the small PSS that is seen in the $N\!\!=\!\!7$ exact calculations [Fig.\ \ref{fig2}(b) and \ref{fig2}(d)] since we have not included any extrinsic pulse errors.  
The dashed reference shown in Fig. \ref{fig2}(a)-\ref{fig2}(d) is the $N\!\!=\!\!7$ exact calculation of CPMG for ${\Omega_{z}}=0$, with ${\mathcal H}_{0}$ set to zero during the pulses \cite{RelTheory}.  Given the strength of the pulses, $\frac{\omega_{1}}{2\pi\Gamma}= 108$, we were surprised to see any difference in the simulations.

\begin{table}[b]
\caption{\label{tab:table2} 
Toggling frame Hamiltonians $\tilde{\mathcal H}(t_{i})$ during each interval $i$ for the CPMG cycle $\{\tau$-$180_{Y}$-$2\tau$-$180_{Y}$-$\tau\}$ with pulse time $t_p$.
Here, $\mathcal H_{yy}\!\!=\!\!\sum_{j>i}^{N} B_{ij}\Big(3I_{y_{i}}I_{y_{j}}\!-\!\vec{I}_{i}\cdot\vec{I}_{j}\Big)$, ${\mathcal H}^{A}_{Y}\!\!=\!\!\sum_{j>i}^{N}\frac{3}{2}B_{i j}(I_{x_i}I_{z_j}\!+\! I_{z_i}I_{x_j})$, and ${\mathcal H}^{S}_{Y}\!\!=\!\!\sum_{j>i}^{N}\frac{3}{2}B_{i j}(I_{z_i}I_{z_j}\!-\!I_{x_i}I_{x_j})$.
The factors ($C_{\theta_{i}}$, $S_{\theta_{i}}$, $C_{2\theta_{i}}$, $S_{2\theta_{i}}$)$\equiv$($\cos{\theta_{i}}$, $\sin{\theta_{i}}$, $\cos{2\theta_i}$, $\sin{2\theta_i}$) 
are time-dependent, as $\theta_{i}\equiv\omega_{1}$$t_{i}$, and $0\leq t_{i}\leq T_{i}$ \cite{mehringText, AHT, ernstText}.}
\begin{ruledtabular}
\begin{tabular}{lll}
$i$ \hfill	& $\: T_{i}$ \hfill		& \hspace{1in} $\tilde{\mathcal H}(t_{i})$\\
\hline
1 \hfill& $\: \tau$  	\hfill& \hspace{2cm} $\!+\Omega_{z}I_{z_{T}}\!+\!{\mathcal H}_{zz}$ \\
2 \hfill& $\: t_{p}$  	\hfill&  $\!+\Omega_{z}(I_{z_{T}}C_{\theta_{i}}\!+\!I_{x_{T}}S_{\theta_{i}})\!-\!\frac{1}{2}{\mathcal H_{y y}}\!+\!{\mathcal H}^{S}_{Y}C_{2\theta_{i}}\!+\!{\mathcal H}^{A}_{Y}S_{2\theta_{i}}$ \\
3 \hfill& $\: 2\tau$  	\hfill&  \hspace{2cm} $\!-\Omega_{z}I_{z_{T}}\!+\!{\mathcal H}_{zz}$ \\
4 \hfill& $\: t_{p}$  	\hfill&  $\!-\Omega_{z}(I_{z_{T}}C_{\theta_{i}}\!+\!I_{x_{T}}S_{\theta_{i}})\!-\!\frac{1}{2}{\mathcal H_{y y}}\!+\!{\mathcal H}^{S}_{Y}C_{2\theta_{i}}\!+\!{\mathcal H}^{A}_{Y}S_{2\theta_{i}}$ \\
5 \hfill& $\: \tau$  	\hfill&  \hspace{2cm} $\!+\Omega_{z}I_{z_{T}}\!+\!{\mathcal H}_{zz}$ \\
\end{tabular}
\end{ruledtabular}
\end{table}

To better understand the origin of the PSS in the simulations, we applied average Hamiltonian theory to CPMG with nonzero pulse duration (the other sequences will be treated elsewhere \cite{RelExpt,RelTheory}). 
The toggling frame Hamiltonian $\tilde{\mathcal H}(t_{i})$ in Table \ref{tab:table2} is used to calculate the leading terms of the
average Hamiltonian $(\bar{\mathcal H}=\bar{\mathcal H}^{(0)}+\bar{\mathcal H}^{(1)}+\ldots )$ \cite{mehringText, AHT, ernstText}.
The zeroth order term is
\begin{equation}
\label{CPMG0avham}
\bar{\mathcal H}^{(0)}_{\mathrm{CPMG}}=\frac{1}{t_c}\big(4\tau {\mathcal H_{z z}} - t_p {\mathcal H_{y y}}\big),
\end{equation}
where the cycle time $t_c = 4\tau + 2 t_p $.
The time-dependent terms within each pulse (see Table \ref{tab:table2}) give rise to a first order term:
\begin{eqnarray}
\label{CPMG1avham}
\bar{\mathcal H}_{\mathrm{CPMG}}^{(1)} &=& \frac{-i}{2 t_c \hbar}\frac{t_p}{\pi} \Big\{ t_p [\mathcal H_y^A, \mathcal H_y^S + \mathcal H_{yy}]\nonumber\\
&&+\:(8\tau + 2t_p)[\Omega_z I_{x_T}, \Omega_z I_{z_T} + \mathcal H_{yy}]\Big\}.\;\;\;\;\;\;\;
\end{eqnarray}
In contrast, the full $\bar{\mathcal H}$ for CPMG with $\delta$-function $\pi$ pulses is simply ${\mathcal H_{zz}}$.  
Since the experimental consequences of finite pulses are not obvious from Eqs. (\ref{CPMG0avham}) and (\ref{CPMG1avham}), we used simulations to study their effects.
Focusing on the simplest case of ${\Omega_{z}}=0$ leaves only one commutator in Eq. (\ref{CPMG1avham}). 

\begin{figure}[b]
\includegraphics[width=3.4 in]{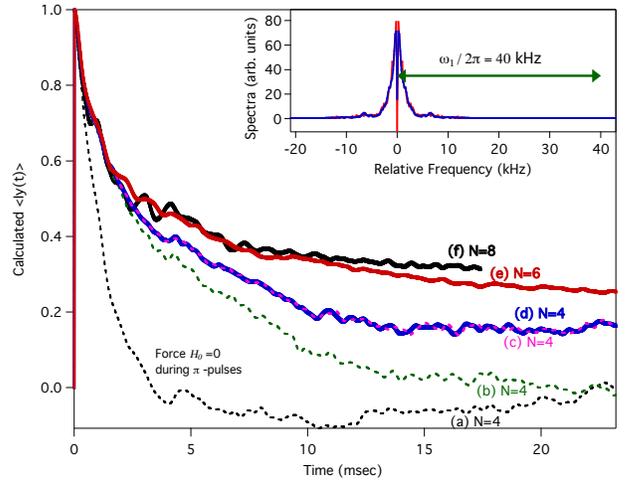}
\caption{\label{fig3}(color online).
CPMG calculations for pure dipolar decay (${\Omega_{z}}=0$, $\frac{\omega_1}{2 \pi}\!\!=\!\!40$ kHz, $\tau$$=\!\!1 \mu$s). Each curve averages 400 DRs [exception (f): 80 DRs] of $N$ spins on a
silicon lattice, with $\gamma'\!\!=\!\!5\gamma$ of $^{29}$Si. 
Several approximations are used for $N\!\!=\!\!4$: (a) setting ${\mathcal H}_{0}=0$ during pulses,
(b) using $\bar{\mathcal H}^{(0)}_{CPMG}$ only, and
(c) using $\bar{\mathcal H}^{(0)}_{CPMG}+\bar{\mathcal H}^{(1)}_{CPMG}$ only.  
Exact calculations for  (d) $N\!\!=\!\!4$, (e) $N\!\!=\!\!6$, and (f) $N\!\!=\!\!8$ show that the tail height depends on $N$ (even N are compared to 
avoid artifacts \cite{walls}).  Inset shows the pulse strength used ($\omega_{1}\gg 2\pi\Gamma$) compared to  the calculated spectra for $N\!\!=\!\!4$ (red) and $N\!\!=\!\!6$ (blue).} 
\end{figure}

Figure \ref{fig3} shows calculations of the CPMG sequence that yield a long tail with parameters ${\Omega_{z}}\!\!=\!\!0$, $\frac{\omega_1}{2 \pi}\!\!=\!\!40$ kHz, and $\tau\!=\!1$ $\mu$s.  
Setting ${\mathcal H}_{0}$ to zero during the $\pi$ pulses yields a leading term $\frac{4\tau}{t_c}{\mathcal H_{z z}}$, causing the fastest signal decay [Fig.\ \ref{fig3}(a)].
Keeping ${\mathcal H}_{0}$ during each pulse modifies
this leading term to $\bar{\mathcal H}^{(0)}_{\mathrm{CPMG}}$ slowing the signal decay [Fig.\ \ref{fig3}(b)].  Adding in the higher order
correction ($\bar{\mathcal H}^{(1)}_{\mathrm{CPMG}}$) 
slows the decay even more, resulting in the long tail [Fig.\ \ref{fig3}(c)].  This is true even for 
${\Omega_{z}}\neq0$ \cite{RelExpt}.

Average Hamiltonian theory was first used to design line-narrowing sequences by generating particular leading terms in the average Hamiltonian.  For example, pulse sequences can be designed to set $\bar{\mathcal H}^{(0)}\!\!=\!\!0$, which results in a lack of signal decay.  Higher-order corrections $\bar{\mathcal H}^{(1)}\!\!\neq\!\!0$ would then modify this behavior by causing some decay in the signal \cite{mehringText, AHT, ernstText}.  In contrast, Fig.\ \ref{fig3} shows that for the CPMG sequence, higher order corrections to the zeroth order Average Hamiltonian can slow decay and even cause a long tail to develop.

The exact calculations in Fig. \ref{fig3}(d)-\ref{fig3}(f) look very similar at early times, as expected for 
the conditions of the simulations [Fig. \ref{fig3}(inset)].  Most surprisingly, the tail height grows with $N$ [Fig. \ref{fig3}(d)-\ref{fig3}(f)].  Thus, knowing the linewidth or $B_{ij}$ scale is not enough to
predict the shape of the whole curve.  Extrapolating from these results \cite{RelExpt,RelTheory}, it appears that the small $N$ and $B_{ij}$ used in Fig. \ref{fig2}(b) inhibit any tail.  Moreover, a noticeable tail should emerge in Fig. \ref{fig2}(b) if the simulations could use $N>20$.

\begin{figure}
\includegraphics[width=3.4 in]{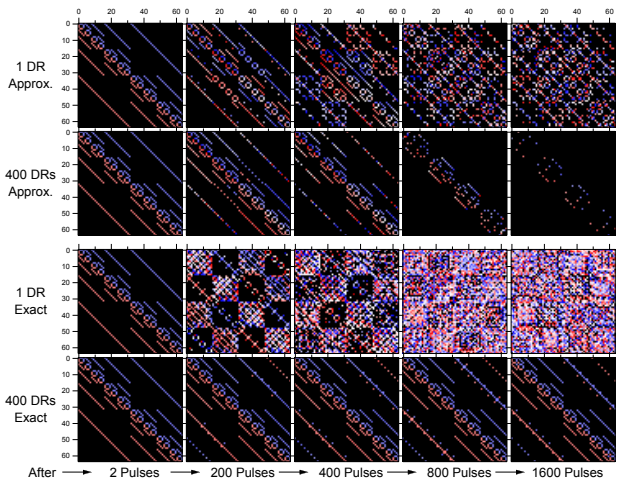}
\caption{\label{fig4}(color online).
Calculated snapshots of a 6 spin ($2^{6}\times2^{6}$) density matrix  evolving during CPMG with conditions as in Fig. \ref{fig3}.
The red-white-blue color scale shows the phase angle; black cells have negligible magnitude.
The top two rows use the $\delta$-function pulse approximation. The bottom two rows use exact $\pi$ pulses. \cite{epaps}
} 
\end{figure}

While the long-lived coherence in CPMG is reminiscent of spin locking [3], the underlying dynamics caused by strong $\pi$ pulses induce important time-dependent changes in the system.  More information is revealed by visualizing the entire time-evolved density matrix as shown in Fig.\ \ref{fig4}.
After two strong $\pi$ pulses (Fig.\ \ref{fig4}, leftmost column) the density matrix still looks like the initial state $\rho(0^+) = I_{y_{T}}$.
The top two rows use the $\delta$-function pulse approximation.  
In each DR (first row), the flow of coherence is restricted to the $\pm1$-quantum coherence cells \cite{slichter, ernstText}, consistent with the selection rules for pure dipolar coupling and $\delta$-function $\pi$ pulses.  
Over $400$ DRs (second row), all quantum coherences average to zero after many pulses, so the measurable signal decays as in Fig.\ \ref{fig3}(a). 
The bottom two rows use exact $\pi$ pulses [see Eq.\ (\ref{Hp})].  
In each DR (third row), coherence flows throughout the entire density matrix, since exact $\pi$ pulses [see Eqs.\ (\ref{CPMG0avham}) and (\ref{CPMG1avham})] open many coherence transfer pathways \cite{ernstText}.  
Despite this complexity, observable coherences emerge after averaging over $400$ DRs (fourth row),  even after many pulses, a counterintuitive result that yields the 
smooth curve in Fig.\ \ref{fig3}(e).

Nonzero pulse duration has been studied for line-narrowing sequences built around $\pi/2$ pulses, and it is not considered to be an error \cite{mehringText}.
For example, the leading term of the average Hamiltonian for the Ostroff-Waugh experiment \cite{ostroffwaugh} is identical for both instantaneous and nonzero duration $\pi/2$ pulses,
$\bar{\mathcal H}^{(0)}_{OsWa}\!=\!-\frac{1}{2}\mathcal H_{yy}$.  
In contrast, nonzero duration $\pi$ pulses introduce completely new operators ($\neq\!\!\mathcal{H}_{zz}$) 
into the average Hamiltonian of CPMG [Eqs.\ (\ref{CPMG0avham}) and (\ref{CPMG1avham})].  Nonzero duration effects should be maximal for $m\times\pi$ pulses, where $m=1,2,3,\ldots$\:.

In general, related effects could arise whenever the applied pulse term does not commute with the spin-spin interaction (e.g., Ising or general anisotropic Heisenberg couplings). 
Examples include ESR experiments on dilute moments
or bang-bang control sequences \cite{Chitty-Chitty} applied to systems of qubits with weak always-on coupling.
Local pulses that address only a subset of coupled spins are also susceptible to these effects.  
Developing an improved understanding of these many-body corrections to pulse action 
will enable the rational design of pulse sequences optimized to boost signal-to-noise, narrow spectra,
and achieve desired coherence transfer pathways.

We thank J. Murray, K. MacLean, X. Wu, E. Paulson, and K.W. Zilm for their experimental assistance and C.P. Slichter, V.V. Dobrovitski, S.M. Girvin, J.D. Walls, and M. M. Maricq for helpful discussions. 
Silicon samples were provided by R. Falster (MEMC) and T.P. Ma.
This work was supported in part by the National Security Agency (NSA) and Advanced Research and Development Activity (ARDA)
under Army Research Office (ARO) contracts No. DAAD19-01-1-0507 and No. DAAD19-02-1-0203, by the NSF ITR Program under grant No. DMR-0342157, and also by NSF grant No. DMR-0207539.


\begin{references}
\bibitem{slichter}C. P. Slichter, {\em Principles of Magnetic Resonance} (Springer, New York, 1990), 3rd ed.
\bibitem{abragam}A. Abragam, {\em The Principles of Nuclear Magnetism} (Oxford University Press, Oxford, 1983), 2nd ed.
\bibitem{mehringText}M. Mehring, {\em Principles of High Resolution NMR in Solids} 
    (Springer-Verlag, Berlin, 1983), 2nd ed.
\bibitem{ernstText}R. R. Ernst, G. Bodenhausen, and A. Wokaun, {\em Principles of Nuclear Magnetic Resonance in One and Two Dimensions} 
    (Clarendon, Oxford, 1987).
\bibitem{siYale} A. E. Dementyev, D. Li, K. MacLean, and S. E. Barrett, Phys. Rev. B {\bf 68}, 153302 (2003).
\bibitem{hahn}E. L. Hahn, Phys. Rev. {\bf 80}, 580 (1950).
\bibitem{siJapan}S. Watanabe and S. Sasaki, Jpn. J. Appl. Phys. {\bf 42}, L1350 (2003).
\bibitem{siStanford} T. D. Ladd, D. Maryenko, Y. Yamamoto, E. Abe, and K. M. Itoh, Phys. Rev. B {\bf 71}, 014401 (2005).
\bibitem{AHT}U. Haeberlen and J. S. Waugh, Phys. Rev. {\bf 175}, 453 (1968).    
\bibitem{cpmg}H. Y. Carr and E. M. Purcell, Phys. Rev. {\bf 94}, 630 (1954); S. Meiboom and D. Gill, Rev. Sci. Instrum. {\bf 29}, 688 (1958).
\bibitem{spinningC60} C. S. Yannoni, R. D. Johnson, G. Meijer, D. S. Bethune, and J. R. Salem J. Phys. Chem. {\bf 95}, 9 (1991); R. Tycko, R. C. Haddon, G. Dabbagh, S. H. Glarum, D. C. Douglass, and A. M. Mujsce, J. Phys. Chem. {\bf 95}, 518 (1991).
\bibitem{levstein} Similar C$_{60}$ data have been reported by: M. B. Franzoni and P. R. Levstein, Phys. Rev. B {\bf 72}, 235410 (2005). 
\bibitem{RelExpt}Dale Li {\em et al.}, arXiv:0704.3620.
\bibitem{BB1}S. Wimperis, J. Magn. Reson. Ser. A {\bf 109}, 221 (1994).
\bibitem{slava} V. V. Dobrovitski (private communication).
\bibitem{RelTheory}Yanqun Dong {\em et al.} (unpublished).
\bibitem{ostroffwaugh} E. D. Ostroff and J. S. Waugh,  Phys. Rev. Lett. {\bf 16}, 1097 (1966).
\bibitem{Chitty-Chitty}L. Viola and S. Lloyd, Phys. Rev. A {\bf 58}, 2733 (1998).
\bibitem{walls}J. D. Walls and Y.Y. Lin, Solid State Nucl. Magn. Reson., {\bf 29}, 22 (2006).
\bibitem{epaps}See EPAPS Doc. No. E-PRLTAO-98-001716 to view the movies underlying Fig.\ 4. For more information on EPAPS, see http://www.aip.org/pubservs/epaps.html.

\end{references}
\end{document}